\documentclass[conference]{IEEEtran}
\IEEEoverridecommandlockouts
\usepackage{amsmath,amssymb,amsfonts}
\usepackage{algorithmic}
\usepackage{graphicx}
\usepackage{tabularx}
\usepackage{textcomp}
\usepackage{xcolor}
\usepackage{comment}
\def\BibTeX{{\rm B\kern-.05em{\sc i\kern-.025em b}\kern-.08em
    T\kern-.1667em\lower.7ex\hbox{E}\kern-.125emX}}

\usepackage[noadjust]{cite}

\newcommand{\ignore}[1]{}
\newif\ifsubmit
\submittrue

\ifsubmit
    \newcommand{\tekin}[1]{}
    \newcommand{\zliu}[1]{}
    \newcommand{\ian}[1]{}
    \newcommand{\ziling}[1]{}
    \newcommand{\yunhuiz}[1]{}
\else
    \definecolor{gray}{rgb}{0.66, 0.66, 0.66}
    \definecolor{dgreen}{rgb}{0.48, 0.62, 0.21}
    \definecolor{dblue}{rgb}{0.00, 0.00, 0.75}
    \definecolor{dred}{rgb}{0.67, 0.22, 0.22}
    \newcommand{\tekin}[1]{{\color{dblue}[Tekin: #1]}}
    
    \newcommand{\zliu}[1]{[{\textcolor{blue}{ Zhengchun: #1 }}]}
    \newcommand{\ziling}[1]{[{\textcolor{dred}{ Ziling: #1 }}]}
    \newcommand{\yunhuiz}[1]{[{\textcolor{dgreen}{ Yunhui: #1 }}]}
    \newcommand{\ian}[1]{[{\textcolor{red}{ Ian: #1 }}]}
\fi

\begin{document}
\bstctlcite{IEEEexample:BSTcontrol}

\title{Deep Learning-based Low-dose Tomography Reconstruction with Hybrid-dose Measurements}


\author{
    \IEEEauthorblockN{
    Ziling Wu\IEEEauthorrefmark{1}, 
    Tekin Bicer\IEEEauthorrefmark{2}\IEEEauthorrefmark{3}, 
    Zhengchun Liu\IEEEauthorrefmark{2}, 
    Vincent De Andrade\IEEEauthorrefmark{3}, 
    Yunhui Zhu\IEEEauthorrefmark{1},  
    Ian T. Foster\IEEEauthorrefmark{2}
    }
    \IEEEauthorblockA{\IEEEauthorrefmark{1}Bradley Department of Electrical \& Computer Engineering, Virginia Polytechnic Institute and State University
    \\\{zilingwu, yunhuiz\}@vt.edu}
    \IEEEauthorblockA{\IEEEauthorrefmark{2}Data Science \& Learning Division and \IEEEauthorrefmark{3}X-Ray Science Division, Argonne National Laboratory
    \\\{tbicer, zhengchun.liu, vdeandrade, foster\}@anl.gov}
}

\maketitle

\begin{abstract}
Synchrotron-based X-ray computed tomography is widely used for investigating inner structures of specimens at high spatial resolutions. 
However, potential beam damage to samples often limits the X-ray exposure during tomography experiments. 
Proposed strategies for eliminating beam damage 
also decrease reconstruction quality.
Here we present a deep learning-based method to enhance low-dose tomography reconstruction via a hybrid-dose acquisition strategy composed of extremely sparse-view normal-dose projections and full-view low-dose projections.
Corresponding image pairs are extracted from low-/normal-dose projections to train a deep convolutional neural network, which is then applied to enhance full-view noisy low-dose projections.
Evaluation on two experimental datasets under different hybrid-dose acquisition conditions show significantly improved structural details and reduced noise levels compared to uniformly distributed acquisitions with the same number of total dosage.
The resulting reconstructions also preserve more structural information than reconstructions processed with traditional analytical and regularization-based iterative reconstruction methods from uniform acquisitions. 
Our performance comparisons show that our implementation, HDrec, can perform denoising of a real-world experimental data 410x faster than the state-of-the-art Xlearn method while providing better quality.  
This framework can be applied to other tomographic or scanning based X-ray imaging techniques for enhanced analysis of dose-sensitive samples and has great potential for studying fast dynamic processes.
\end{abstract}


\section{Introduction}
\begin{figure*}[t]
\centering
\includegraphics[width = 0.8\linewidth]{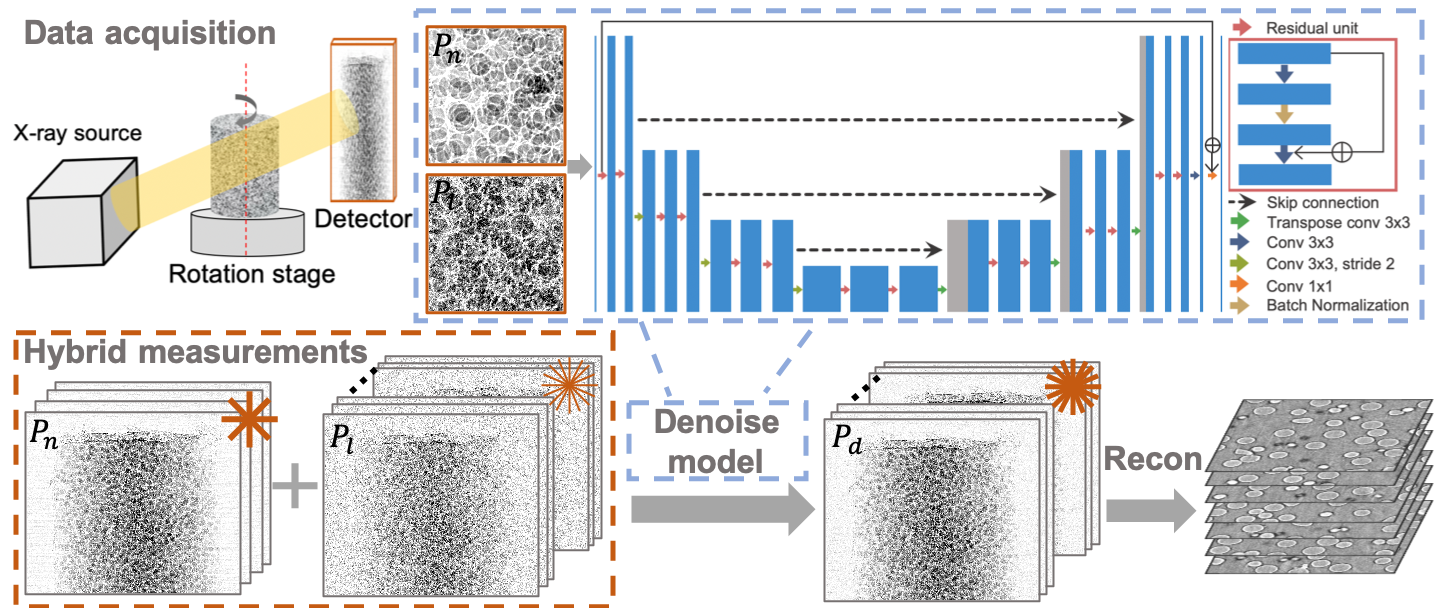}
\caption{The architecture of the proposed deep learning reconstruction framework for low-dose tomography under hybrid-dose acquisition mode (HRrec). Hybrid-dose projections are recorded and corresponding image pairs are extracted to learn the denoising model. Learned model is then applied to full-view low-dose projections to obtain denoised full-view projections $P_d$. Finally, image slices are retrieved with traditional reconstruction algorithms.}
\label{fig1}
\end{figure*}

X-ray computed tomography (CT) is widely used at synchrotron radiation facilities~\cite{SedighRahimabadi2020ReviewCharacterization}, for example to detect defects in structural materials~\cite{Wu2019AutomaticMeasurements}, understand pore formation in shales~\cite{Kanitpanyacharoen2013ASLS}, and investigate network constructions in porous materials~\cite{Yang2020QuantitativeMethodology}. 
During X-ray tomographic acquisition, a detector records the decreased photon flux after X-ray photons penetrate a target from multiple orientations: see Fig.~\ref{fig1}. 
Each recorded image is called a \emph{projection}. 
A specialized algorithm is then used to reconstruct the distribution of X-ray attenuation in the volume being imaged, providing internal three-dimensional (3D) morphology at high spatial and temporal resolutions. 
However, radiation dosage during data acquisition is a limiting factor for most tomography experiments, as extended X-ray exposure can cause beam damage to the sample. 
Reducing the number of projection views (sparse-view) or the X-ray tube current can be used to limit beam exposure (low-dose), but low-dose data acquisition schemes yield noisy measurements that significantly reduce the quality of the reconstructed image.

Many algorithms have been proposed to improve the reconstruction quality for low-dose tomography. Deep learning, particularly deep neural network (DNN), based reconstruction methods have shown particular promises. 
Existing DNN-based approaches fall into three main groups depending on the operation domains: 1) right after the data acquisition to denoise noisy raw measurements (\emph{measurement domain learning}\/)~\cite{Lee2018Deep-neural-networkReconstruction,Yang2018Low-doseNetwork}, 2) during tomographic reconstructions to represent retrieval algorithms with learning-based methods (\emph{model-based approaches}\/)~\cite{Chen2018LEARN:CT,Wurfl2018DeepProblems}, and  3) after tomographic reconstructions to denoise noisy images retrieved with conventional reconstruction algorithms (\emph{image domain learning}\/)~\cite{Jin2017DeepImaging,Pelt2018AAnalysis,Liu2020TomoGANTomography,WuRobustNetworks,liu2019deep}.
Almost all developed methods for low-dose CT reconstruction improvement require access to full high-quality ground truth data. 
However, there are limitations to acquire these data, since high-quality data typically require high-dose data acquisition that may not be possible given experimental constraints on X-ray exposure, as is the case in dose-sensitive samples and dynamic processes. 
Recently proposed unsupervised learning processes have achieved good results~\cite{Wang2020Semi-supervisedRestoration}. 
However, these approaches require estimation of an additive noise model to correct the data. Further, because a learned model cannot easily be applied to different samples, a new model must be trained whenever a new sample or feature is encountered.

We explore here a new approach in which signals from low-dose projections are enhanced during the acquisition itself, as shown in Fig.~\ref{fig1}.  In this approach, several low-dose projections are collected with their corresponding normal-dose counterparts. These low-/normal-dose image pairs are used to learn a mapping from features and noise through a DNN-based learning network; the resulting model which then be used to enhance other noisy low-dose projections. 
This hybrid-dose data acquisition strategy requires few normal-dose projections, which in turn accelerates acquisitions, minimizes dose effects on samples, and also simplifies the transfer learning.
Specifically, we make the following contributions in this paper:
\begin{itemize}
  \item We develop a state-of-the-art DNN-based denoising approach in the \emph{measurement domain} to achieve high-quality reconstructions. Our method learns the denoising model from \emph{extremely sparse-view normal-dose projections} and \emph{full-view low-dose projections}\zliu{I guess here should be ``their corresponding low-dose measurement"}. The trained model can be used to enhance and denoise full-view low-dose projections.    
  \item We explore the strategy to distribute dosage smartly with best reconstruction quality. The denoising and reconstruction results are obtained by models learned with different number of low-/normal-dose projection pairs and different dosage value for low-dose projections.
  \item We evaluate our method on two real-world experimental datasets, and demonstrate that the method can provide excellent projection denoising and reconstruction results. The results also outperform traditional regularization-based reconstruction and deep learning-based denoising approaches in terms of image quality and computational efficiency.
\end{itemize}
We show that the combination of deep learning with hybrid-dose acquisition can enable high-quality tomographic reconstructions with low radiation dose. 
Our method can be applied to other tomographic or scanning based X-ray imaging techniques and has great potential for studying fast dynamic processes. 

\section{Background and Related work}
We briefly explain the tomographic data acquisition and reconstruction process, and then review methods used to improve noisy tomographic reconstructions.  

\subsection{Physical models and CT reconstruction}
In X-ray imaging, X-rays penetrate a sample and the incident energy, as attenuated by the density and the thickness of the sample, is recorded: see Fig.~\ref{fig1}. This measurement results in an X-ray image called a \emph{projection}. 
This energy decrease is characterized by Beer's Law, which describes intensity reduction as a function of X-ray energy, path length, and material linear attenuation coefficient with $I=I_0e^{-\mu L}$, 
where $I_0$ and $I$ are the initial and final X-ray intensity, $\mu$ is the material's linear attenuation coefficient (units 1/length) and $L$ is the length of the X-ray path.
During a tomographic data acquisition, many projections are collected from different orientations so that the sample is imaged from every angle and fully covered according to the requirement of the {\em Nyquist sampling} theorem. 
Typically, each projection is exposed to X-rays (radiation dose) for a fixed amount of time. 
The forward model of CT acquisition is finalized with inverse equation $P = Ax$, where $P \in \mathbb{R}^{m\cdot p}$ denotes the projection data, $A \in \mathbb{R}^{n\times m\cdot p}$ denotes the forward projection model, $x \in \mathbb{R}^n$ denotes the reconstructed image, $m$ the number of pixels in one projection, $p$ the number of projections, and $n$ the number of pixels/voxels in the reconstructed image. 

The tomographic reconstruction is the process of recovering 3D volume of the sample $x$ from a set of 2D projections $P$.
An analytical solution of tomographic reconstruction problem is calculated by solving inverse equation directly.
However, the reconstruction quality is highly sensitive to the experimental setup, number of projections, the quality of acquired projections, and etc.
For instance, short-exposure data acquisitions yield noisy reconstructed images as too few photons can result in extremely noisy projections and thus corrupted reconstructions.

Regularization-based iterative reconstruction methods are proposed to improve reconstruction quality. 
These methods aim to optimize an objective function by considering the physical model in the iterative model. 
Many iterative reconstruction methods are developed based on different regularization terms, include dictionary learning~\cite{Xu2012Low-doseLearning}, nonlocal means~\cite{Chen2009BayesianPrior}, total variance and variances~\cite{Sidky2008ImageMinimization,Liu2012Adaptive-weightedReconstruction}. 
While iterative methods provide improved reconstruction quality under these undesired conditions, they are significantly more computationally demanding compared to analytical methods~\cite{hidayetoglu2020peta, hidayetouglu2019memxct, bicer2017trace, bicer2015europar}. 

\subsection{Deep learning-based modeling for noise reduction}
Deep learning techniques, particularly deep neural network (DNN), have been actively developed in recent years, and applied to various applications.
In particular, deep learning provides new thinking and tremendous potential in the field of tomographic image reconstruction \cite{Wang2018ImageLearning}. 
Many algorithms have been developed to improve reconstruction quality with DNN-based techniques, which are mainly grouped as \emph{measurement domain learning}\/, \emph{model-based approaches}\/, and \emph{image domain learning}\/, as discussed before. 

Traditionally, denoising the reconstructed images is widely used as the sample features are easier to detect and learn compared to projections that are difficult to distinguish with overlapping features. 
Many DNN-based approaches have been developed to denoise reconstructed images from traditional analytical method, such as multi-resolution convolutional neural network (CNN)~\cite{Jin2017DeepImaging}, mixed-scale dense CNN~\cite{Pelt2018AAnalysis}, generative adversarial networks~\cite{Liu2020TomoGANTomography}, and hierarchical synthesis-based CNN~\cite{WuRobustNetworks}.
However, denoising the reconstructed images requires training a model with two 3D volumes from the same sample: one from noisy or incomplete projections and the other from normal-dose projections. 
In addition, transferring a trained model to another sample is not trivial, since the learned features and the scale of the noise are likely to have mismatches, e.g., a model trained with shale sample cannot work on a plant cell sample.
Another issue with denoising reconstructed images is that training a model with reconstructed images and applying it to other samples runs the risk of introducing new artifacts in the images or losing structural information~\cite{Dong2019AData}. 

Denoising methods have been used within~\cite{Chen2018LEARN:CT} or as a replacement for~\cite{Wurfl2018DeepProblems} tomographic reconstruction algorithms, with promising results.
However, model-based denoising methods are computationally expensive, with costs similar to those of the regularization-based iterative reconstruction methods mentioned in the previous section. 

Denoising measurement data directly before reconstruction has also been proposed. 
Similar to image domain learning, the DNN-based denoising model is trained with noisy and clean image pairs. 
For example, Lee et al.~\cite{Lee2018Deep-neural-networkReconstruction} proposed DNN-based algorithms to synthesize missing-view data with full-view measurements. 
Yang et al.~\cite{Yang2018Low-doseNetwork} presented a DNN-based method trained using low-/high-dose projection pairs to enhance low-dose X-ray tomographic projections. \ian{Previous sentence unclear.}\zliu{revised }
\ziling{I delete this sentence here. It seems too many points said for Yang's work.}

Overall, the signal-to-noise ratio decreases as photon counts drop. The structural information about the sample cannot be reconstructed successfully unless regularization methods or denoising methods are used. While there are ample studies on reconstructed image denoising, researches on denoising projections are limited. 
\section{Method}
In this section, we explain our DNN-based denoising method via a hybrid-dose acquisition scheme to improve low-dose tomography reconstructions.

\subsection{Overview: Feature extraction with hybrid-dose acquisition}
Fig.~\ref{fig1} shows the proposed DNN-based reconstruction framework for low-dose tomography under hybrid-dose acquisition scheme, which is referred to as HRrec. 
This framework consists of three parts. 
First, hybrid-dose projections are recorded including \emph{several extremely sparse normal-dose projections ($P_n$)} and \emph{full-view low-dose projections ($P_l$)}. 
The stars on the top right corner of sub-figures show different acquisition conditions, where the number of total star lines represents the number of total projections and the thickness of each star line represents the dosage value for each projection. Then corresponding image pairs are extracted from low-/normal-dose projections to train the network. The architecture of the DNN is discussed in the next section. The trained DNN denoising model was then used to enhance the full-view low-dose projections ($P_l$). 
Finally, a traditional reconstruction algorithm, such as filtered back-propagation, is adopted to retrieve images with the denoised projections ($P_d$). 

\subsection{DNN architecture}
Our denoising model improves the state-of-the-art U-Net architecture~\cite{Ronneberger2015U-net:Segmentation} with residual blocks~\cite{He2016DeepRecognition}, referred to as {\em residual U-net}, to facilitate the information flow.
As shown in Fig.~\ref{fig1}, the inputs to the neural networks are the corresponding image pairs extracted from low-/normal-dose projection pairs.
The network consists of two principal parts: the image {\em encoder} and {\em decoder}. 
The encoder uses four successive encoding processes to extract features, each containing two residual units, as shown on the top right legend. This use of residual units is proved to be efficient to learn noise models for image denoising in computer vision and computed tomography applications \cite{Chen2017Low-DoseNetwork}. 

The encoded features then pass through three decoding processes to decode the features. 
Each process is the same as before with two residual units.
At the end, a convolution layer with one 1$\times$1 kernel generates the single channel image to match the target image. 
Feature maps generated from transposed convolution layers are concatenated with the preceding feature maps of the same scale from the encoder part.

\subsection{Objective function}
In this section, we present the loss functions used in the residual U-Net, which are the main components to quantify the difference between the noisy input and ground truth. 

1) {\bf \emph{$\boldsymbol{\ell_1}$ loss}}
is a mean-based metric that encourages the pixels of the output image $y$ to match exactly the pixels of the target image $\hat{y}$\zliu{well, it actually tries to match the median of target}. It is similar to the $\ell_2$ loss (mean-squared error),
but does not over-penalize larger errors between a denoised image and the ground truth as does $\ell_2$ loss; this prevents over-smoothness and blurring.
The $\ell_1$ loss also helps further improve signal-noise-ratio.
In our low-dose CT image denoising task, the $\ell_1$ loss function is: 
\begin{equation}
\ell_1(\hat{y}, y) = \frac{1}{C W H}\lVert \hat{y}-y\rVert_1,
\end{equation}
where $\hat{y}$ and $y$ are the ground truth (normal-dose projection) and a denoised projection, respectively. $C$, $W$, and $H$ are the image width, height, and depth. 

2) {\bf \emph{Perceptual loss}} helps to retain texture and structural details of the denoised image by comparing high level differences, like content and style discrepancies, between images. 
We implement it by calculating the mean squared error of features extracted by a pre-trained VGG network \cite{Simonyan2015VeryRecognition}:
\begin{equation}
l_{VGG}(\hat{y}, y) = \frac{1}{C_f W_f H_f}\lVert VGG(\hat{y})-VGG(y)\rVert_2, 
\end{equation}
where $C_f, W_f, H_f$ represent the dimension of the feature maps extracted by the pre-trained $VGG$ network.  \\

\noindent
{\bf \emph{Overall loss function}}:
We combine {\em $\ell_1$ loss} to ensure pixel identity and {\em perceptual loss} to keep high-level texture and structural details: 
\begin{equation}
l = \alpha \ell_1+\beta l_{VGG}
\end{equation}
where the  coefficients $\alpha$ and $\beta$ balance these two loss terms, which are decided empirically. 
In the training stage, the total loss between the output $y$ and a normal-dose projection $\hat{y}$ was calculated for each step and then back-propagated for the neural network optimization. 
By combining these two loss functions, we insure the quality of refined projection image.
\subsection{Model training}
Our residual U-Net with hybrid loss was trained using image patches with the size of 128$\times$128 extracted from low-/normal-dose projections pairs, then applied on entire full-view low-dose projections. 
We show different number of projection pairs to explore the relationship between the denoised results and the requirement of normal-dose projections in the result section. 
For each case, 80\% of total low-/normal-dose projections pairs are used for training and the rest are used for validation. We use  the adaptive momentum estimation (Adam) to optimize our residual U-Net with the batch size of 16. 
We set the learning rate to 1$\times$10$^{-4}$. 
We implemented the network with Tensorflow.

\subsection{Tomographic reconstruction}
We feed the full-view low-dose projection images $P_l$ into the trained denoising model, as shown in Fig.~\ref{fig1}, producing full-view high quality denoised projection images $P_d$. 
We then perform tomographic reconstruction with the open-source TomoPy toolbox \cite{Gursoy2014TomoPy:Data}, using
the Fourier grid algorithm (GridRec) with a Parzen filter to balance reconstruction speed and accuracy. 
Other tomographic reconstruction methods can also be used. 
\section{Evaluation}

\begin{figure}[h]
\centering
\includegraphics[width = 0.85\linewidth]{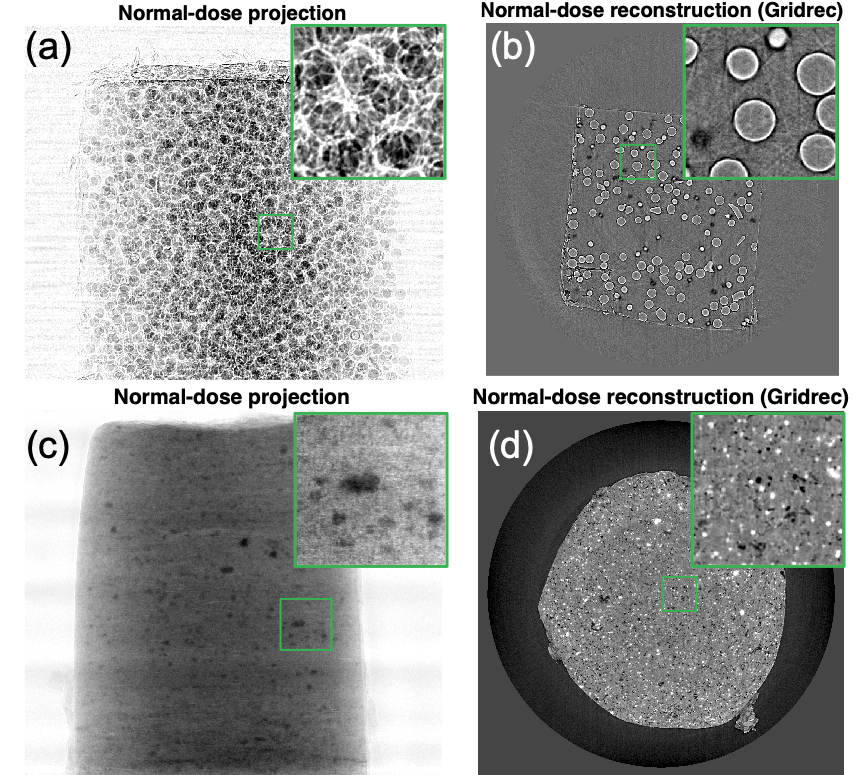}
\caption{Overview of the glass and shale samples. One normalized projection slice for the (a) glass and (c) shale samples. One reconstructed image slice with Gridrec algorithm for the (b) glass and (d) shale sample.}
\label{fig2}
\end{figure}

We now present the denoising and reconstruction performance of our proposed method when applied to the glass and shale datasets (Fig. \ref{fig2}) from TomoBank \cite{DeCarlo2018TomoBank:Science} under different configurations.
We used one 
NVIDIA Tesla V100 SXM2 (32GB memory) GPU card for training, denoising, and reconstruction. 

We evaluate our proposed methods against a variety of data acquisition schemes and compare our method to the state-of-the-art learning-based algorithm, Xlearn \cite{Yang2018Low-doseNetwork}, and one iterative total variation-based (TV) regularization reconstruction \cite{Sidky2008ImageMinimization} method both in terms of image quality and computational time requirements. 

\subsection{Glass sample}
\begin{figure}[h]
\centering
\includegraphics[width = 0.85\linewidth]{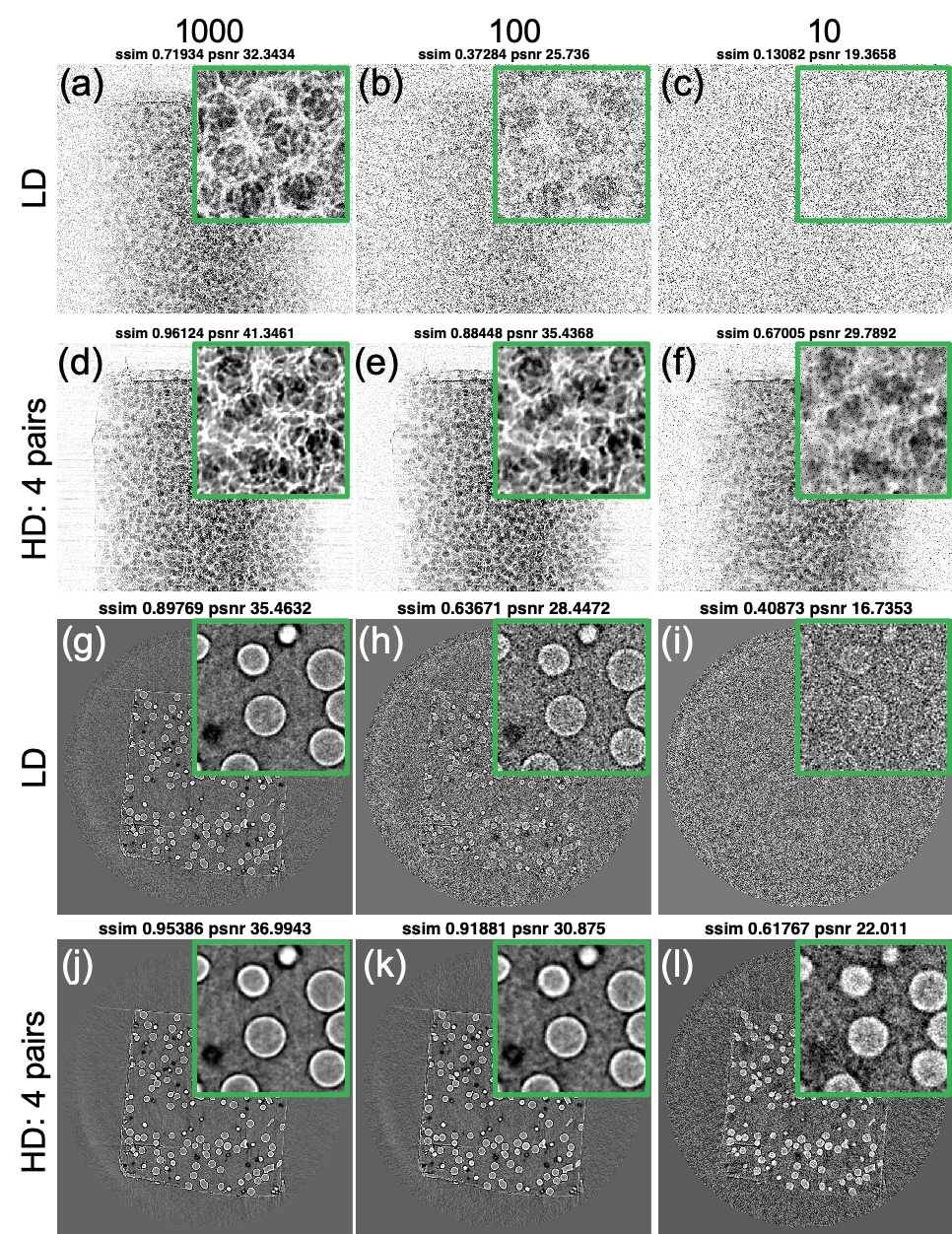}
\caption{Simulated low-dose projections and denoised projections and reconstruction results for the glass sample. The first and third rows show the simulated low-dose projections and reconstruction results with the dosage value of (a, g) 1000, (b, h) 100 and (c, i) 10 per projection. The second and last rows show the denoised projections and reconstruction results enhanced with the models trained with four projection pairs and the dosage value of (d, j) 1000, (e, k) 100 and (f, l) 10 in the low-dose simulations. \tekin{LD refers to low-dose whereas HD represents the denoised projections with hybrid-dose and their reconstructions. We use only 4 normal-dose projections to train and enhance low-dose configurations.}
}
\label{fig3}
\end{figure}

The glass sample contains 20\% volume fractions of borosilicate glass spheres encased in a polypropylene matrix.
It was measured at the 2-BM fast tomography beamline of the Advanced Photon Source (APS), Argonne National Laboratory (ANL). 
The experiment was performed with source energy of 27.4 keV and exposure time of around 0.0001s as the normal-dose projections $P_n$. 
1500 projections were taken over 180$^{\circ}$. 
Fig.~\ref{fig2}(a) shows one normalized projection in the normal-dose measurements. Multiple bubbles are summed along the X-ray propagation path for each ray path, resulting in the structure overlap. The green box on the top right corner shows the zoom-in structural details. 
Fig.~\ref{fig2}(b) shows one reconstructed image slice retrieved from full-view normal-dose measurements with Gridrec algorithm. The reconstructed structures are seen clearly in this figure.  

In order to simulate low-dose projections from these normal-dose measurements $P_n$, we used Siddon’s ray-driven forward projection method \cite{Siddon1985FastArray}, and added Poisson noise as: 
\begin{equation}
P_l \sim \textrm{Poisson}\{b_0 P_n\}/b_0
\label{Pos_sim}
\end{equation}
where $b_0$ is the blank scan factor representing the dosage value per ray/pixel and $P_l$ is the simulated low-dose detector measurements. No electronic readout noise was simulated. The dosage reduction ratio of simulated low-dose measurements can be adjusted by setting the number of photons per projection for the blank scan factor $b_0$. In the glass sample, the average number of photons for normal-dose projections per pixel is measured to be around 5000. For simplification, we use 5000 to represent the number of photons received per projection for each normal-dose projection.\zliu{I feel the simplification does not sound right for X-ray scientist.}\ziling{I know. Could we keep this way now? I am actually not very clear about the dosage distribution for each projection.}
Low-dose projections are simulated with Eq.~\ref{Pos_sim} by setting $b_0$ = 1000, 100, and 10 as lower dosage values compared to the dosage of normal projections. 
We use the Structural Similarity Index (SSIM) and Peak signal-to-noise ratio (PSNR) to quantify the structural similarity and noise level compared to original normal-dose projections.
As shown in Fig.~\ref{fig3}(a--c), the SSIM of simulated projections decreases with decreasing dosage per projection, and consequently the structural details become more difficult to distinguish.

Denoising models are trained with simulated low-dose and experimentally measured normal-dose projection pairs to enhance the full-view low-dose projections $P_l$ and thus obtain the denoised projections $P_d$.
Fig.~\ref{fig3}(d--f) show the denoised projection results $P_d$ with four different low-/normal-dose projection pairs at 0$^{\circ}$, 45$^{\circ}$, 90$^{\circ}$, and 135$^{\circ}$ with the dosage value of low-dose projections $b_0$ equals to 1000, 100, and 10. 
Denoised projections $P_d$ (Fig.\ \ref{fig3}(d--f)) demonstrate significant noise reduction and structural details improvement compared to corresponding low-dose projections $P_n$ (Fig. ~\ref{fig3}(a--c)). 
This is also validated with improved SSIM and PSNR values. 
These improvements are even more significant for the projections from lower-dosage cases. 
For the low-dose projections with dosage values of 100 and 10, it is difficult to distinguish the structures of the glass phantoms with very low SSIM and PSNR values.
By using the denoised models 
the denoised projections (Fig.~\ref{fig3}(e, f)) are significantly improved.
\begin{figure}[h]
\centering
\includegraphics[width = 0.85\linewidth]{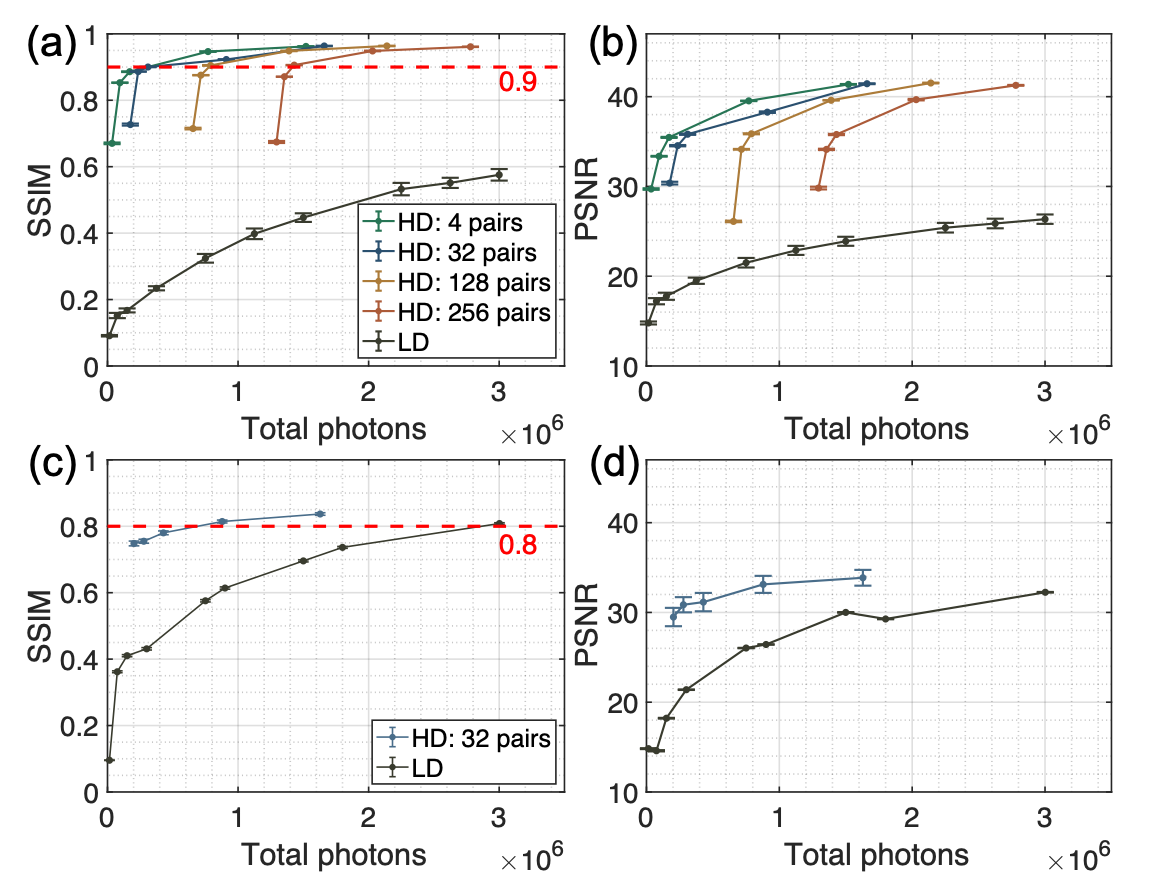}
\caption{SSIM and PSNR distribution with respect to the number of total photons under each configuration for the (a, b) glass and (c, d) shale sample. Green, blue, yellow, and orange lines in (a, b) denotes the performance of enhanced projections with different number of low-/normal-dose projection pairs (4, 32, 128, and 256). Five dots on each line represent different dosage values $b_0$ = 10, 50, 100, 500, 1000 for corresponding low-dose projections. Blue line in (c, d) denotes the performance of enhanced projections with the models trained with 32 low-/normal-dose projection pairs. Five dots on each line represent different dosage values $b_0$ = 50, 100, 200, 500, 1000 for low-dose projections. Black line represents the performance of pure low-dose projections.}
\label{fig4}
\end{figure}

We also compare the denoised projection performance under more hybrid-dose acquisition schemes in terms of SSIM and PSNR values to explore the influence of different number of low-/normal-dose projection pairs. 
Fig.~\ref{fig4}(a, b) depict SSIM and PSNR values in terms of mean and standard variation for the denoised projections enhanced with 4, 32, 128, and 256 low-/normal-dose projection pairs, shown in green, blue, yellow, and orange lines, respectively. 
Five dots on each line represent the dosage value of low-dose projections $b_0$, which are 500, 100, 50, and 10 respectively.
We calculate the total photons for each configuration as x label by summing up all required sparse-view normal-dose and full-view low-dose projections. 
By comparing these four lines under different hybrid-dose acquisition modes, we could conclude that more projection pairs don't improve enhanced projection performance. 
On the contrary, the least low-/normal-dose projection pairs, which is four projection pairs in our study, always performs the best in terms of SSIM and PSNR values. 
We also calculate SSIM and PSNR values of low-dose projections under uniformly distributed acquisition modes. Ten different dosage values per projection are simulated with respect to the total photons and treat as the baseline, shown as the black line in figure ~\ref{fig4}(a, b). 
The performance of non-hybrid measurements (black line) is always worse than that of the denoised projections with the network enhancement (green, blue, yellow and orange lines) under the fixed number of total photons. 
In other words, when the number of total photons is limited, uniformly distributing the total photons to each projection \zliu{I suppose this is the current state of the art strategy using convention reconstruction method. if so, worth mention it} \ziling{Here is comparing projection performance. There is no reconstruction involved here. I will mention the reconstruction method when comparing reconstruction results in the next paragraph.} 
can results in mediocre or sub-optimal projection quality.
As empirically shown, it is better to concentrate the photons to several projections as normal-dose projections, and uniformly distribute the rest as low-dose projections. \tekin{reworded}


\tekin{please check the following and see if it makes sense}
Next, we study the reconstruction results from full-view normal-dose projections $P_n$, which can also be noisy due to the insufficient dose, full-view low-dose projections $P_l$. 
Fig.~\ref{fig2}(b) shows the the tomographic reconstructions from  $P_n$, whereas Fig.~\ref{fig3}(g--i) and (j--l) illustrate reconstructions from $P_l$ and their corresponding {\em denoised} versions, respectively.
\zliu{why?}
\ziling{Explained in this way now. }
The SSIM value of the final reconstruction slice for the dosage value of 1000 (Fig.~\ref{fig3}(j)) improves from 0.90 to 0.95. 
For the reconstruction results with pure low-dose measurements with dosage of 100 (Fig.~\ref{fig3}(h)) and 10 (Fig.~\ref{fig3}(i)), it is difficult to distinguish the structures of the glass phantoms with very low SSIM values.
The final reconstructions (Fig.~\ref{fig3}(k, l)) from enhanced projections with corresponding trained models are improved significantly in terms of visual quality and quantitative SSIM value.  
\begin{figure}[h]
\centering
\includegraphics[width = 0.85\linewidth]{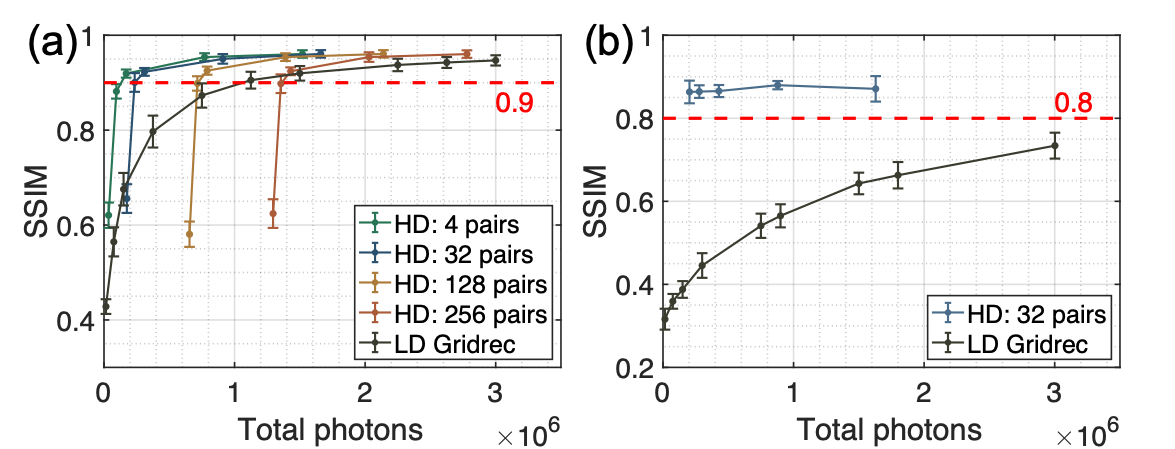}
\caption{SSIM distribution of reconstructed slices with respect to the number of total photons under different hybrid-dose acquisition schemes for the (a) glass and (b) shale sample. Green, blue, yellow, and orange lines in (a) denotes the performance of enhanced projections with four different number of low-/normal-dose projection pairs (4, 32, 128, and 256). Five dots on each line represent different dosage values $b_0 =$ 10, 50, 100, 500, and 1000 for low-dose projections. Blue line in (b) denotes the performance of enhanced projections with the models trained with 32 projection pairs. Five dots on each line represent different dosage values $b_0 = $ 50, 100, 200, 500, and 1000 for low-dose projections. The black line represents the performance of pure low-dose projections.}
\label{fig6}
\end{figure}
\begin{figure}[h]
\centering
\includegraphics[width = 0.85\linewidth]{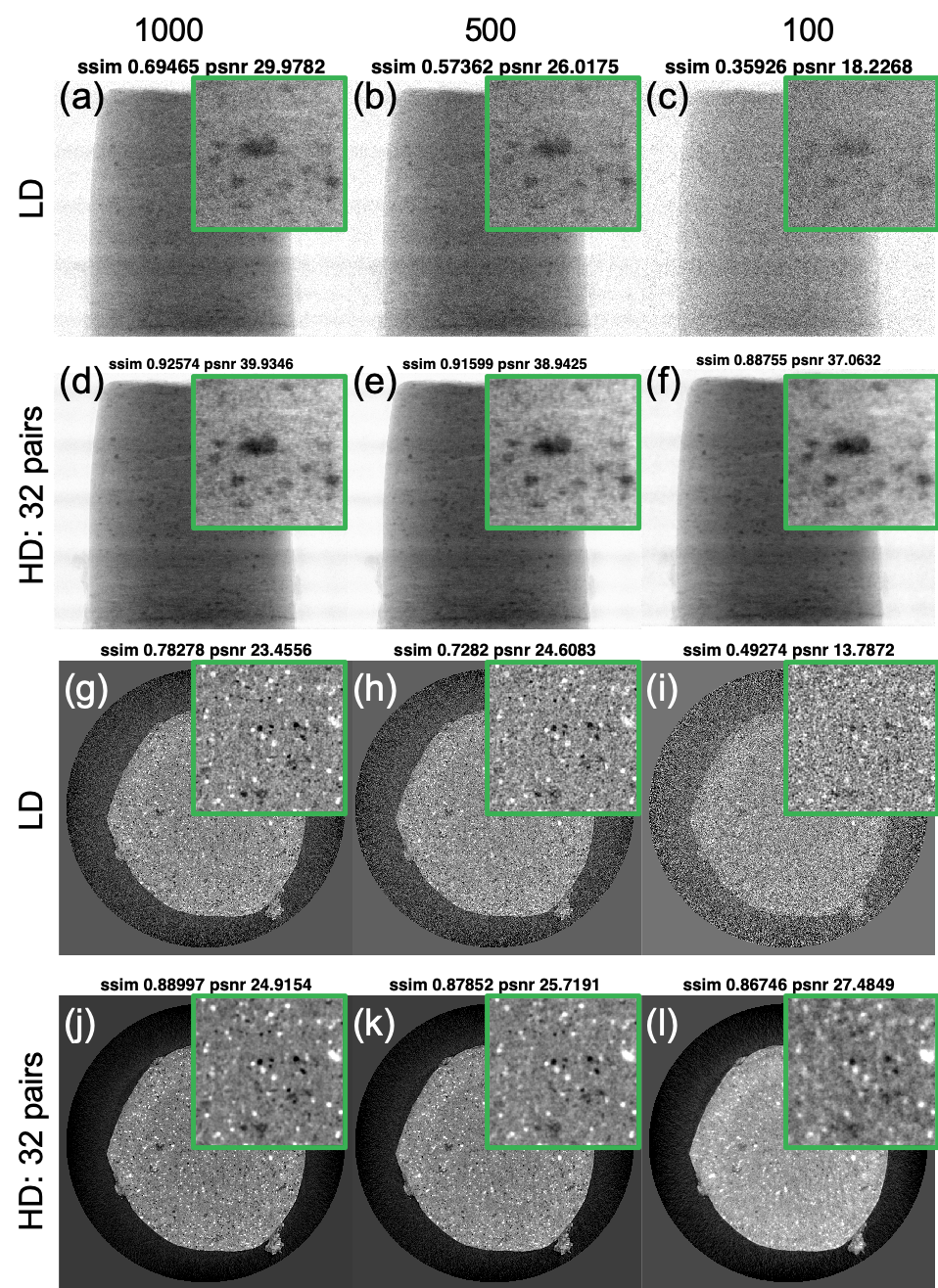}
\caption{Simulated low-dose projections and denoised projections and reconstruction results for the shale sample. The first and third rows show the simulated low-dose projections with the dosage value of (a, g) 1000, (b, h) 500 and (c, i) 10 per projection. The second and last rows show the denoised projections enhanced with the models trained with 4 low-dose and normal-dose projection pairs and the dosage value of (d, j) 1000, (e, k) 500 and (f, i) 100 for the low-dose simulations.}
\label{fig7}
\end{figure}

We also compare the mean SSIM values of the final reconstruction volume performance to explore the influence of different number of normal-dose and low-dose projection pairs and how to distribute the total X-ray photons.
We could tell there are some differences between Fig.~\ref{fig4}(a). 
Final reconstruction performed from the learned denoising models with several normal-dose projections outperform uniformly distributed the total photons to each projection. 
The two points beneath the black line are reconstruction results from the models trained with 128 and 256 low-/normal-dose projection pairs and low-dose projection dosage values $b_0$ of 10 and 50.
When we compare the SSIM performance under the fixed number of total photons, the points on green line and blue line significantly outperform these two configuration and also better than the black line which is uniformly distributing dosage for non-hybrid measurements.
These points on the green and blue lines represent reconstruction results trained with 4 and 32 low-/normal-dose projection pairs and low-dose projection dosages $b_0$ of around 500 and 1000. 
This is because if the dosage value of projections is too low, the structural details are smeared by the noise.
Simply adding more normal-dose projections doesn't improve the structural restoration.
In other words, when the number of total photons is fixed, it is better to distribute the photons to fewer projections, and uniformly distribute the rest with not too low dosage value for each projection.
\subsection{Shale sample}
We further validated the proposed method on a second experimental dataset, with quite different structural features, the shale sample. 
Shale is a challenging material because of its multi-phase composition, small grain size, low but significant amount of porosity and strong shape- and lattice-preferred orientation.
In this work, we use a shale sample dataset from the Upper Barnett Formation in Texas. 
We used 1501 projections acquired over 180$^{\circ}$ at APS as the normal-dose projections.
Fig.~\ref{fig2}(c) shows one normalized projection in the normal-dose measurements. 
Multi-phase compositions are summed up along the X-ray propagation path for each ray path, resulting in the structure overlap. 
Fig.~\ref{fig2}(d) shows one reconstructed image slice retrieved from full-view normal-dose measurements with Gridrec algorithm. 
The average number of photons for normal-dose projections per pixel is measured around 4000. 
We simulate the low-dose projections from these normal-dose measurements according to Eqn.~\ref{Pos_sim}. 

Figs \ref{fig7}(a--c) and \ref{fig7}(g--i) show simulated low-dose projections and corresponding reconstructed image slices for low-dose projections with blank scan factor $b_0$ = 1000, 500, and 100. 
Grain structures are smeared by the noise with the decreased dosage value per projection in the low-dose measurements, which is also validated with the SSIM and PSNR of simulated projections and SSIM of reconstructed image slice. 
\begin{figure}[h]
\centering
\includegraphics[width = 0.85\linewidth]{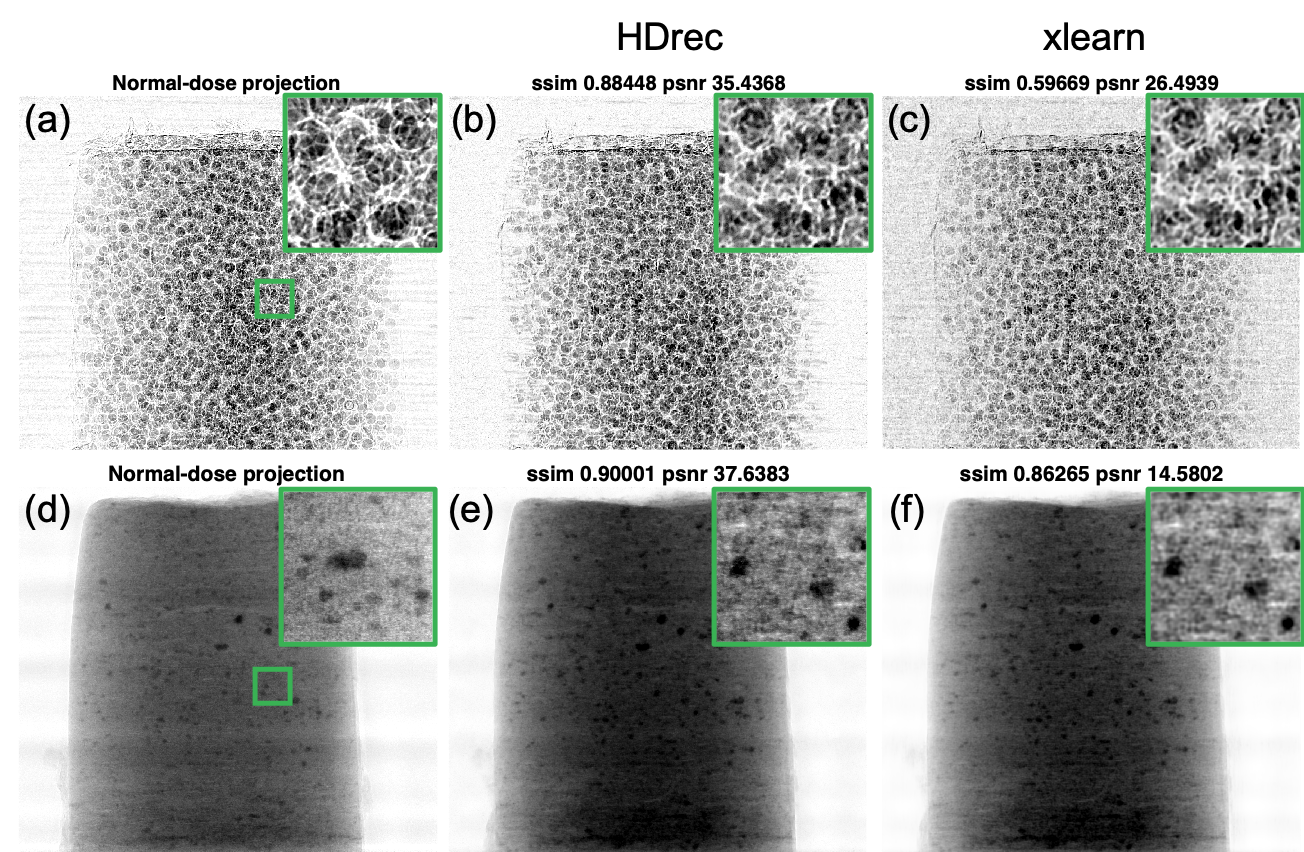}
\caption{Projection denoising performance comparison with Xlearn for the (a--c) glass and (d--f) shale sample: (a, d) normal-dose projections, (b, e) denoised projections with our method, HDrec; (c, f) denoised projections with Xlearn. }
\label{fig11}
\end{figure}
Similar to the glass sample, we trained the residual U-Net to estimate the high-resolution measurements ($P_n$) when provided with full-view low-dose projections $P_l$ ($b_0$ = 1000, 500, 10) with 32 low-/normal-dose projection pairs.
The second row of Fig.~\ref{fig7}(d--f) and Fig.~\ref{fig7}(j--l) show the denoised projections $P_d$ and reconstruction results enhanced by the denoised models.
Denoised projections $P_d$ could remove the noise significantly and improve the structural details compared to corresponding low-dose  projections $P_l$, shown in Fig.\ \ref{fig7}(a--c). 
Tomographic reconstructions (Fig.~\ref{fig7}(j--l)) of corresponding denoised full-view projections also show improved structural restoration, which is also validated with SSIM values. 

We also plot the SSIM and PSNR values of projection slices (Fig.~\ref{fig4}(c, d)) and SSIM values of reconstructed slices (Fig.~\ref{fig6}(b)) vs.\ number of total photons for each configuration. 
The performance of uniform distributed low-dose measurements (black line) is always worse than that of the denoised projections with network enhancement (blue line) under a fixed number of total photons, which is consistent with the glass sample results. 
It is better to concentrate photons to several projections as normal-dose projections, as shown before, and uniformly distribute the rest as low-dose projections.

\subsection{Projection denoising performance comparison with Xlearn}
We compare the denoised projection results against a previously proposed learning-based method, Xlearn. This comparison method was originally trained with several low-/high-dose projection pairs and applied to whole low-dose measurements. For fair comparison with our case, we use the same network architecture and training strategies as Xlearn and retrain for the low-/normal-dose projection datasets used in this work. As shown in fig.~\ref{fig11}, our method achieves better structural restoration and noise removal performance compared to Xlearn.
\begin{figure}[h]
\centering
\includegraphics[width = 0.85\linewidth]{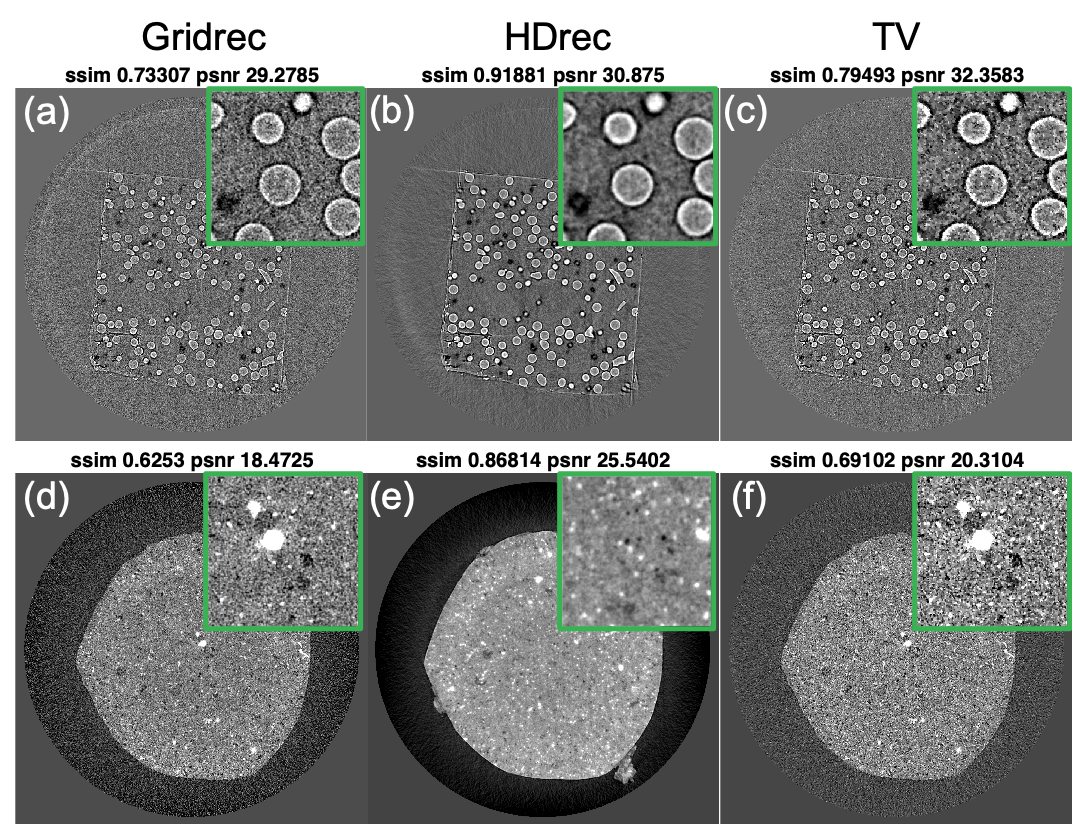}
\caption{Reconstruction performance comparison with total variation-based regularization method for the (a--c) glass and (d--f) shale sample. (a, d) Reconstruction results with Gridrec for low-dose projections of same number of total dosage as hybrid-dose measurements. (b, e) Reconstruction results with Gridrec for denoised projections. (c, f) Reconstruction results with TV-based method for low-dose projections of same number of total dosage as hybrid-dose measurements.}
\label{fig12}
\end{figure}
\subsection{Reconstruction performance comparison with total variation-based regularization method}
We also compare reconstruction image quality with that obtained via the iterative total variation-based (TV) regularization reconstruction method. 
We show the reconstructed results (Fig.~\ref{fig12}(b, e)) generated by the denoising model trained with 32 low-/normal-dose projection pairs and low-dose dosage values of 100 and 200, respectively.
For a fair comparison, we simulate the uniform distributed low-dose projections using the same total dosage as for HDrec: 206 and 285 per projection for glass and shale, respectively.
The Gridrec (Fig.~\ref{fig11}(a, d)) and TV-based (Fig.~\ref{fig11}(c, f)) reconstructions for the glass and shale samples are worse than those obtained with HDrec, both qualitatively and quantitatively. 

\subsection{Computational time comparison}
\zliu{as why compare with Xlearn, I think we need to mention that Xlearn is the only literature that study low-dose denoising in projection domain}
We use the glass sample to compare the computational costs of HDrec, the Xlearn-based projection denoising method, and the TV-based iterative reconstruction method.
All methods are run under the same computer configuration.
1) \emph{Single projection denoising}:
HDrec's use of a fully convolutional neural network allows it to complete in
1.34s: 410 times faster than the 550s taken by Xlearn.
\ziling{I think it doesn't make sense to compare this directly added number. Projections needed to be done for the whole volume for image reconstruction.}
2) \emph{Reconsructing a single image}:
For HDrec and Xlearn, this requires the application of gridrec to denoised projections (around 0.86s for a single image); for TV-based iterative reconstruction, only the reconstruction operation is needed to obtain the final image slice, with average time of around 55.79s.
3) \emph{Total time from denoising of projections to reconstruction}: For a complete dataset, it takes around 1 hour, 9.5 days, and 17 hours 
for HDrec, Xlearn, and the TV-based method, respectively,
corresponding to speedups of 214 and 16 for HDrec over 
Xlearn and the TV-based method, respectively.
\ian{Is my rewording of thie paragraph ok?}\zliu{Yes, much conciser, Thanks} 
\section{Conclusion}
We have presented a deep learning-based enhancement method, HDrec, for low-dose tomography using hybrid-dose measurements, which contains \emph{extreme sparse-view normal-dose projections} and \emph{full-view low-dose projections}. 
The denoised projections and reconstructed slices show significant improvement when compared to xlearn-based projection denoising and TV-based reconstruction methods in terms of image quality and computational efficiency. 
In addition, we provide a strategy to distribute dosage smartly with improved reconstruction quality. 
When total dosage is limited, the strategy of fewer normal-dose projections and not too low full-view low-dose measurements greatly outperforms the uniform distribution of the dosage.  

\section*{Acknowledgment}
\yunhuiz{please add the NSF funding for QMR}
\ziling{Added}
This material was partially supported by the U.S. Department of Energy, Office of Science, Advanced Scientific Computing Research, Basic Energy Sciences under Contract DE-AC02-06CH11357 and National Science Foundation under Contract CMMI-1825646.
We gratefully acknowledge the computing resources provided on Bebop, a high-performance computing cluster operated by the Laboratory Computing Resource Center at ANL.

\bibliographystyle{IEEEtran}
\bibliography{bicer,references}

\section*{Government License}
The submitted manuscript has been created by UChicago Argonne, LLC, Operator of Argonne National Laboratory (``Argonne''). Argonne, a U.S.\ Department of Energy Office of Science laboratory, is operated under Contract No.\ DE-AC02-06CH11357. The U.S.\ Government retains for itself, and others acting on its behalf, a paid-up nonexclusive, irrevocable worldwide license in said article to reproduce, prepare derivative works, distribute copies to the public, and perform publicly and display publicly, by or on behalf of the Government.  The Department of Energy will provide public access to these results of federally sponsored research in accordance with the DOE Public Access Plan. http://energy.gov/downloads/doe-public-access-plan.

\end{document}